\documentclass[cits,useAMS]{PoS}
\usepackage{amssymb}
\usepackage{fontenc}
\usepackage{times}
\usepackage{mathptmx}
\usepackage{float}
\usepackage{subfig}
\graphicspath{/misc/home/sd805/LaTeX/Dublin_conf/}

\title{Temporal Studies of Supergiant Fast X-ray Transients}

\ShortTitle{Temporal Studies of Supergiant Fast X-ray Transients}

\author{\speaker{Sebastian Drave}$^{1}$, Antony Bird$^{1}$, David Clark$^{2}$, Vanessa McBride$^{1}$, Adam Hill$^{3}$, Vito Sguera$^{4,6}$, Simone Scaringi$^{5}$, Angela Bazzano$^{6}$, Lee Townsend$^{1}$\\
        $^{1}$School of Physics and Astronomy, University of Southampton, University Road, Southampton, SO17 2RZ, UK \\
	$^{2}$Centre d'Etude Spatiale des Rayonnements, CNRS/UPS, BP 4346, 31028 Toulouse, France \\
	$^{3}$Laboratoire d'Astrophysique de Grenoble, UMR 571 CNRS, Universit$\acute{e}$ Joseph Fourier, BP 53, 38041, Grenoble, France \\
	$^{4}$INAF-IASF, Instituto di Astrofisica Spaziale e Fisica Cosmica, VIA Gobetti 101, 40129 Bologna, Italy\\
	$^{5}$Department of Astrophysics, IMAPP, Radbound University Nijmegen, PO Box 9010, 6500 GL Nijmegen, the Netherlands\\
	$^{6}$INAF-IASF Roma, Instituto di Astrofisica Spaziale e Fisica Cosmica, Via del Fosso del Cavaliere 100, 00133 Roma, Italy\\
        E-mail: \email{sd805@soton.ac.uk}}

\abstract{SFXTs are a new class of HMXB unveiled by INTEGRAL. They are extreme systems characterised by very short outbursts (a few hours) and extreme X-ray luminosity dynamic ranges ($\sim$10$^{4}$). Ten confirmed systems are currently known and have shown parallels with both Sg-XRBs and Be-XRBs. Temporal studies across all timescales are key to understanding both the place of SFXTs within the HMXB hierarchy and the accretion mechanisms at work within the objects. Here we present the discovery of a new 51.47 $\pm$ 0.02 day orbital period in the SFXT XTE J1739$-$302 using \emph{INTEGRAL} observations. We also present a higher time resolution study of the SFXTs SAX J1818.6-1703, IGR J16479-4514 and IGR J16465-4507 using RXTE that shows newly discovered flaring activity.}

\FullConference{8th INTEGRAL Workshop ``The Restless Gamma-ray Universe''\\
                 September 27-30 2010\\
                 Dublin Castle, Dublin, Ireland}

\begin{document}

\section{Introduction}

Supergiant Fast X-ray Transients (SFXT) are a new class of HMXB that have been unveiled over the lifetime of the \emph{INTEGRAL} \cite{WinklerINTEGRAL} mission. These sources are characterised by very short outbursts \cite{sguera2005} and an association with OB supergiant companions \cite{chaty2008}. There are currently 10 confirmed SFXTs clustered along the Galactic Plane, as well as several candidates: X-ray sources that show the required fast X-ray flaring but as yet have not been associated with supergiant companions. With the determination of source distances from optical/IR spectroscopy \cite{rahoui2008}, peak outburst luminosities of 10$^{36}$ $-$ 10$^{37}$ erg s$^{-1}$ have been deduced. Combined with the observation of quiescence states at 10$^{32}$ erg s$^{-1}$ this illustrates a very high X-ray dynamic range of 10$^{4}$ $-$ 10$^{5}$ within these systems. They show parallels with the two well-studied classes of high mass X-ray binaries (HMXBs), the Supergiant (Sg-XRBs) and Be X-ray binaries (BeXRBs). X-ray pulsations have been detected in four confirmed SFXTs, implying that they are likely accreting neutron stars. 

Current research centers on the deduction and understanding of the orbital parameters and accretion mechanisms within these extreme members of the HMXB family. On a broader scale consideration is also given as to whether the SFXTs constitute a new class of HMXB, or are members of the known classes exhibiting extreme X-ray behaviour which is most suited to detection by wide-field monitoring missions such as \emph{INTEGRAL}. A fundamental building block of this understanding is knowledge of the mass transfer and accretion mechanisms taking place within SFXTs. The models used to describe this process fall into two main categories; the `clumpy wind' and magnetic gating models. In the first instance the rapid, large X-ray outbursts are described by the compact object travelling through the structured stellar wind of the OB supergiant companion: either in the form of a spherically symmetric, `clumped' stellar wind outflow \cite{intZand2005} or via the passage through an enhanced wind density region in the equatorial plane of the companion star \cite{sidoli2007}. Within this construction the difference in emission profiles observed compared to classical wind-fed Sg-XRBs can be explained by variations in the orbital profile of the compact objects within the systems, generally using longer, more eccentric orbits. The `magnetic gating' models, however, utilise varying magnetic and co-rotation radii of the neutron star to impede and release the accretion flow onto the neutron star surface, hence creating the large X-ray dynamic ranges observed \cite{greb2007}. However these models fail to produce the recurrent, periodic outbursts that are becoming an evermore common feature of SFXT emission. This combined with the fact that the neutron star would be required to posses a magnetar strength magnetic field to create the observed outbursts have led to the `clumpy wind' models becoming the more accepted method of SFXT outburst production. Gated accretion scenarios could still have a role to play in SFXT outbursts however as Ducci et al. \cite{Ducci2010} employed such mechanisms, amongst others, within a clumpy wind model to attempt to explain the shorter timescale structure seen within some SFXT outbursts.

Central to the development of our understanding of SFXTs is the use of temporal analysis of both long and short duration data sets to reveal the physical processes at work within these extreme HMXBs. In these proceedings we present analysis of both a long duration \emph{INTEGRAL}/IBIS data set, and shorter RXTE observations revealing new insights into four SFXT systems. 

\section{XTE J1739$-$302 / IGR J17391$-$3021}

Using $\sim$12.4\,Ms of \emph{INTEGRAL}/IBIS \cite{UbertiniIBIS} an orbital period of 51.47$\pm$0.02\,days has been discovered in the SFXT XTE J1739$-$302. Features within the phase-folded lightcurve of this system showed possible indications of an enhanced equatorial density region. If confirmed these features would imply a link towards the BeXRB class of HMXB and suggest that SFXTs are in fact an intermediate class that bridge the gap between the two original classes of HMXB. For a full description of this study see Drave et al.  \cite{me2010}.

\section{SAX J1818.6$-$1703}

SAX J1818.6$-$1703 was first detected by \emph{BeppoSax}/WFC on 1998 March 11 with a peak flux of $\sim$400 mCrab (9$-$25 keV) \cite{intZand2008} showing an outburst that lasted only 0.1\,d. First detected by \emph{INTEGRAL} on 2003 September 9, two similar flares were observed, again reaching $\sim$400 mCrab \cite{grebenev2005}. Following a \emph{Chandra} localisation the companion was confirmed as an OB supergiant \cite{SAXintzand2006} and the system obtained full SFXT status following a non-detection in a 13\,ks \emph{XMM-Newton} observation, illustrating that SAX J1818.6-1703 posses the required large dynamic range to be considered as such \cite{SAXbozzo2008}. Using long term \emph{INTEGRAL}/IBIS lightcurves Bird et al. \cite{SAXbird2008} detected a period of 30$\pm$0.1\,d in the SAX J1818.6$-$1703 system. It was also observed that SAX J1818.6$-$1703 shows a very high recurrence rate: the source was detected in over 50\% of the periastron passages covered by the data set. 

\begin{figure}[h]
 \centering
 \subfloat{\includegraphics[width=0.48\textwidth]{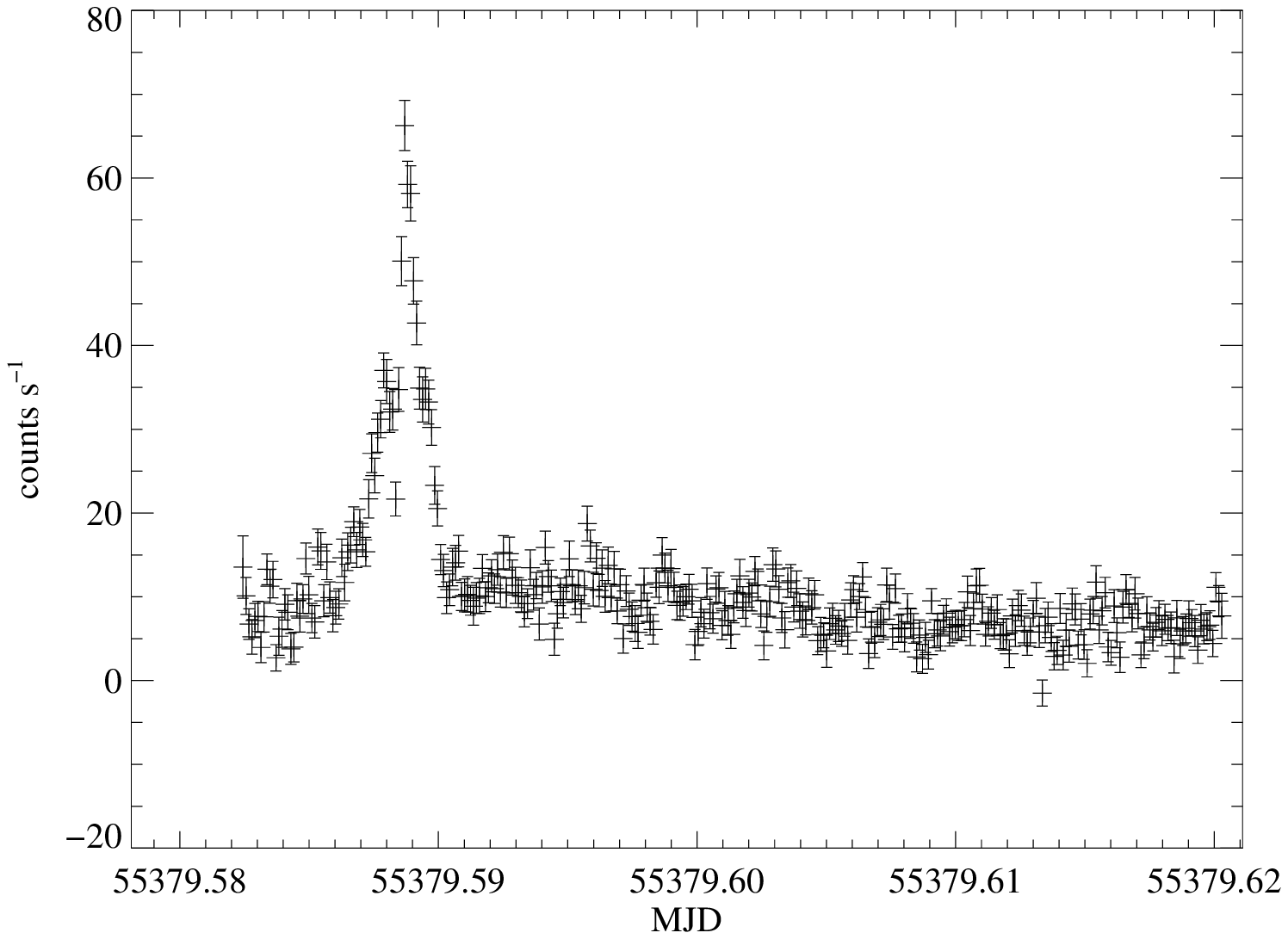}}
 \subfloat{\includegraphics[width=0.48\textwidth]{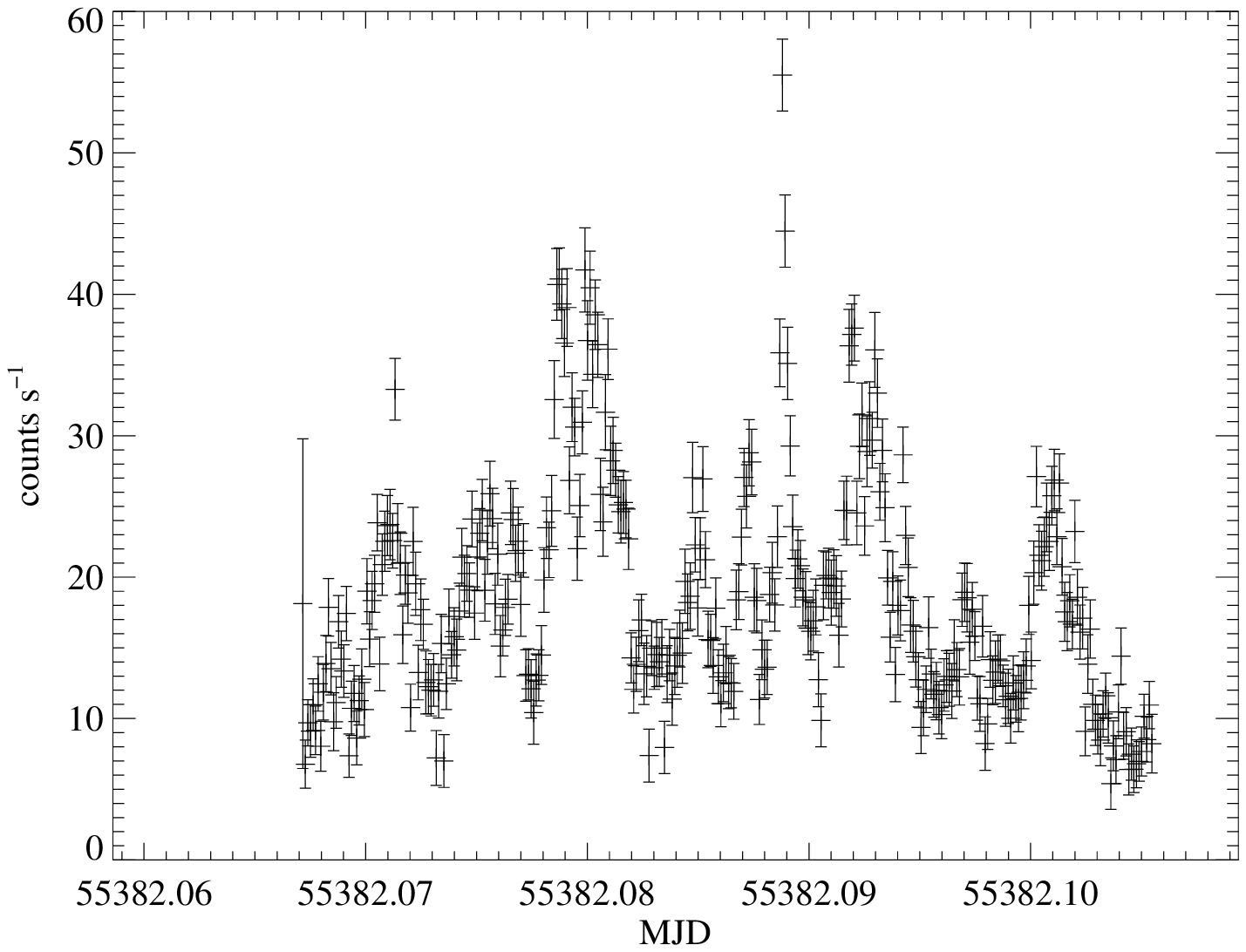}}
 \caption{SAX J1818.6$-$1703 outbursts observed with RXTE showing an isolated, sharp outburst with its peak at MJD\,55379.588 (left) and a longer period of sustained flaring activity with the peak at MJD\,55382.089 (right).}
 \label{SAXJ1818_outbursts}
\end{figure}

Using the orbital ephemeris of Bird et al. \cite{SAXbird2008} 40\,ks of periastron targeted \emph{RXTE}/PCA \cite{RXTEPCA} observations were performed between 2010 July 2 and 2010 July 6 with the aim to study the periastron passage in detail and search for pulsations from the compact object. As expected from the high recurrence rate SAX J1818.6$-$1703 showed a high level of outburst and flaring activity across the periastron passage. Two examples of this activity are shown in Fig. \ref{SAXJ1818_outbursts}. The 10 second binned lightcurves show both singular sharp outbursts (left) and more prolonged periods of high level flaring activity of a comparable flux (right). The large scale variations in outburst profile are likely due to the wind environment and individual wind clump macro properties encountered during each outburst. The finer structure observed in some outbursts could have its origins in more complicated accretion processes. Ducci et al. \cite{Ducci2010} employed the formation and dissipation of transient accretion disks, accretion via the Rayleigh-Taylor instability and X-ray photoionization to explain similar outburst profiles observed in other SFXTs. However due to the long orbit, 30.0\,d, of SAX J1818.6-1703 these mechanisms would require the system to have at least a moderate eccentricity in order to produce the required orbital dynamics in which the observed outburst profiles are able to be created. The unusually high recurrence ratio observed in SAX J1818.6-1703 also suggests that the system may have a significant level of eccentricity. Hence detailed modelling of the outbursts of the SAX J1818.6$-$1703 system could provide constraints on an orbital parameter that is vital for our understanding of SFXTs and the extreme activity they display. Neutron star pulsations were not be detected in any of the observations.

\section{IGR J16479$-$4514 and IGR J16465$-$4507}

IGR J16479-4514 was first reported by Molkov et al. \cite{J16479molkov2003} during an outburst with an average flux of $\sim$12 mCrab (18$-$25 keV) and associated with an O8.5I supergiant companion star at a distance of $\sim$4.9\,kpc by Chaty et al. \cite{chaty2008,rahoui2008}. However a quiescence X-ray luminosity of $\sim$10$^{34}$ erg cm$^{-2}$ s$^{-1}$, measured by Sguera et al. \cite{J16479sguera2008}, is two orders of magnitude higher than that seen in other SFXTs. As a result IGR J16479$-$4514 is considered an intermediate SFXT as it has displayed all aspects of SFXT behaviour but has a higher level of quiescence emission. Jain et al. \cite{J16479jain2009} reported the detection of a 3.32\,d orbital period within the system that also revealed an X-ray eclipse lasting $\sim$0.6\,d. IGR J16465-4507 is also considered an intermediate SFXT, showing a smaller dynamic range than classical SFXTs but also significantly higher than persistent wind-fed HMXBs. First reported by Lutovinov et al. \cite{J16465lutovinov2004} the source produced $\sim$9 and 25 mCrab outbursts on 6 and 7 September 2004 respectively before returning to an undetectable state. An \emph{XMM-Newton} observation a couple of days later identified a supergiant star at $\sim$9.5\,kpc \cite{rahoui2008} as the only bright infrared object within the error circle and detected a 228\,s periodicity within the X-ray lightcurve which was taken as the spin period of the neutron star within the system \cite{J16465lutovinov2005}. The orbital period of the system was reported as 30.32$\pm$0.02\,d by Clark et al. \cite{J16465clark2010}.

\begin{figure}[h]
 \begin{center}
 \includegraphics[scale=0.3]{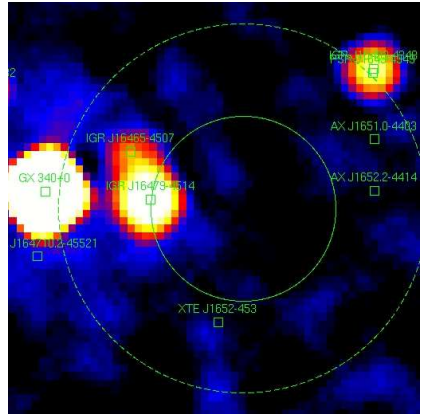}
 \caption{The field around the pointing used for \emph{RXTE}/PCA observations of IGR J16479-4514 and IGR J16465-4507. The solid and dashed circles are the half and zero response contours of the PCA detector respectively. This pointing was chosen to minimise contamination from the bright, persistent source GX 304+0. The image is a survey map from the 4$^{th}$ IBIS/ISGRI catalog \cite{TonyCat4}}
 \label{RXTE_FOV}
 \end{center}
\end{figure}

\begin{figure}[h]
 \centering
 \subfloat{\includegraphics[width=0.48\textwidth]{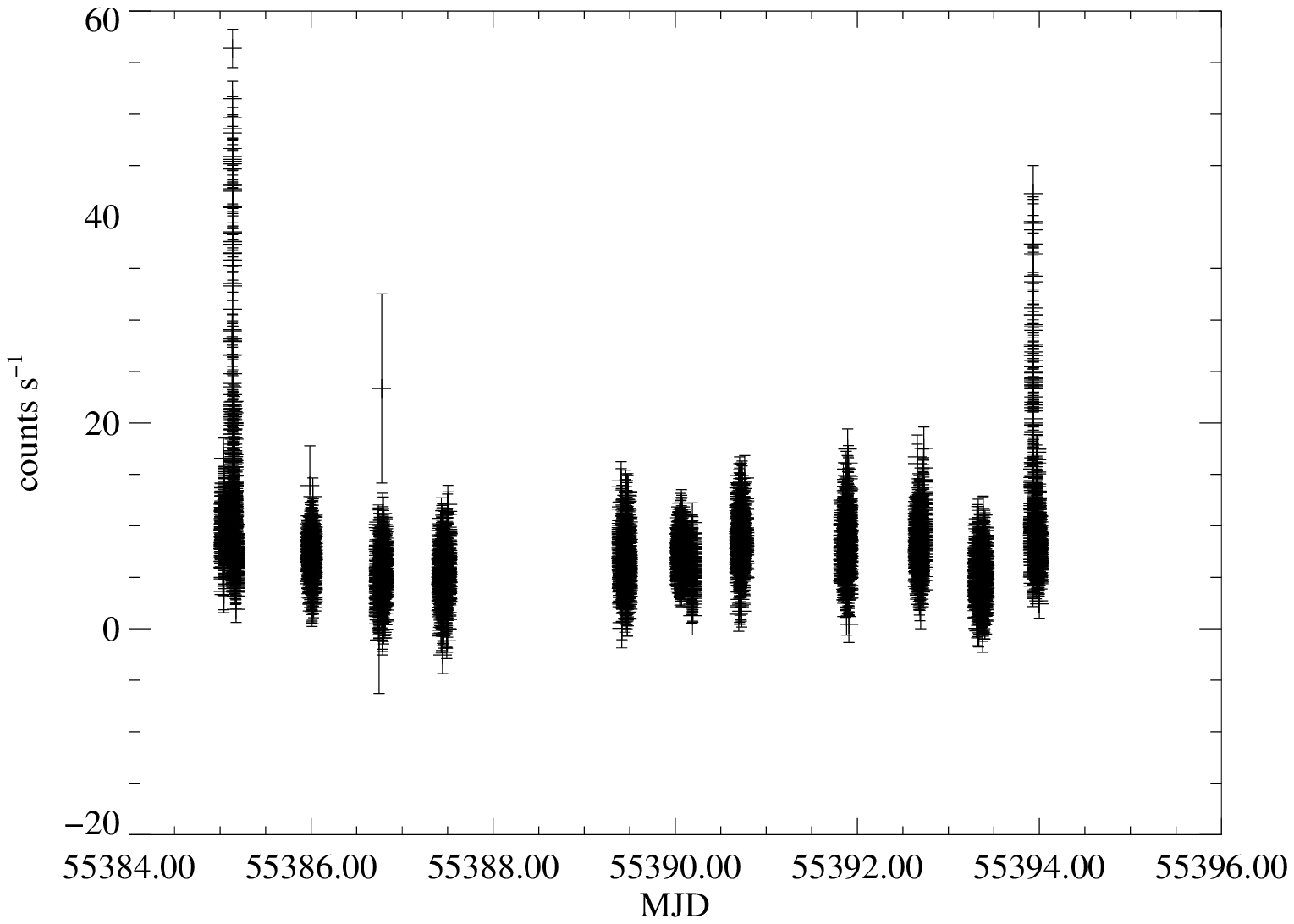}}
 \subfloat{\includegraphics[width=0.48\textwidth]{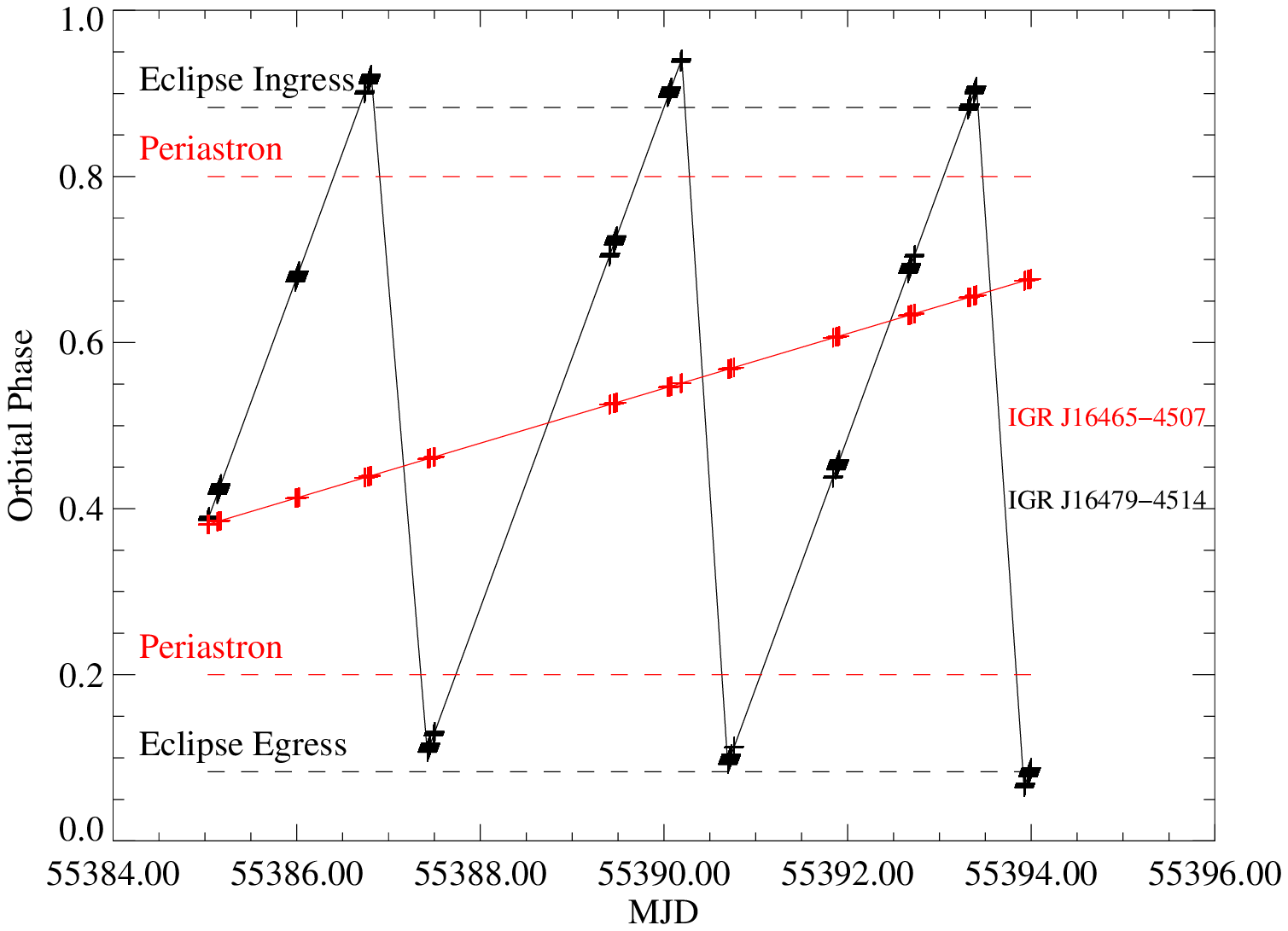}}
 \caption{Left: 10-second binned lightcurve of the \emph{RXTE}/PCA observations showing two distinct outburst events. Right: Observation distribution in IGR J16479$-$4514 (black) and IGR J16465$-$4507 (red) orbital phase. It is observed that the second large outburst occurred whilst IGR J16479$-$4514 was in eclipse.}
 \label{double_bursts}
\end{figure}

Due to the small angular separation of these two sources, 16', a priori knowledge of the orbital periods and the presence of an eclipse in the IGR J16479$-$4514 system was essential in making observations of these systems with \emph{RXTE}/PCA viable. Fig. \ref{RXTE_FOV} shows the field around the two SFXT systems along with the \emph{RXTE}/PCA half and zero response contours (for the pointing used). As can be seen the large FOV of the PCA detector encompasses both of the SFXT systems. Using the known orbital periods and phase location of the X-ray eclipse, 60\,ks of pointed observations were performed covering IGR J16479-4514 whilst out of eclipse, in full eclipse and also during both eclipse ingress and egress. The lightcurve of these observations are shown in Fig. \ref{double_bursts} (left) along with their locations in the orbital phase of both systems (right). As can clearly be seen two outbursts were detected throughout the duration of the observations. 


Inspecting the first outburst it can be seen that it occurred at an IGR J16479-4514 orbital phase of $\sim$0.4, whilst the system was out of eclipse, and at an IGR J16465-4507 orbital phase of $\sim$0.4 also, whilst this system was approaching apastron. As IGR J16465$-$4507 has an orbit ten times longer than that of IGR J16479$-$4514 the neutron star separation at such a phase will be much larger for this system. Hence we interpret the first outburst as being from the IGR J16479$-$4514 system. Upon inspecting the second outburst however it is found that this event occurred whilst IGR J16479$-$4514 was in eclipse. As a result of this we interpret this event to have originated from the IGR J16465$-$4507 system. The spectra of both outburst events were investigated using XSPEC. Using absorbed powerlaw models (PHABS(POWERLAW)) with an additional iron K$\alpha$ emission line the best fits outlined in Table \ref{spectralfits} were achieved. Whilst the spectra are not consistant with one another, the differences do not permit the identification of the source of each outburst. Hence it is only through our knowledge of the orbital characteristics of the sources that such a distinction can be made. 

\begin{table}
 \begin{center}
 \begin{tabular}{c c c c c c}
  \hline
  \multicolumn{1}{c}{Outburst} & \multicolumn{1}{c}{Source} & \multicolumn{1}{c}{$\chi^{2}$/d.o.f.} & \multicolumn{1}{c}{$\Gamma$} & \multicolumn{1}{c}{n$_{H}$} & \multicolumn{1}{c}{Luminosity} \\
  MJD & & & & 10$^{22}$ cm$^{-2}$ & erg s$^{-1}$ (3$-$10 keV) \\ \hline
  55385.136 & IGR J16479$-$4514 & 0.98 & 1.804 & 4.4 & 1.533$\times$10$^{36}$ erg s$^{-1}$ \\
  55393.933 & IGR J16465$-$4507 & 1.30 & 1.501 & 8.5 & 4.076$\times$10$^{36}$ erg s$^{-1}$ \\ \hline
 \end{tabular}
 \caption{Spectral parameters of the observed outbursts using an absorbed powerlaw spectral model.} 
 \label{spectralfits}
 \end{center}
\end{table}

\section{Conclusions}

Timing analysis from the shortest to longest scales provides in-depth diagnostics of the physical processes taking place within SFXT systems. In the above studies we have shown the wealth of different information that can be gathered about these systems, from the 51.47 day orbital period in XTE J1739$-$302 to fast flaring activity which could help constrain the orbital characteristics and accretion processes at work in SAX J1818.6$-$1703. Finally we have shown how long baseline studies with survey missions such as \emph{INTEGRAL} provide the required knowledge to make the studies of systems with small separations possible with instruments to which they previously have been inaccessible. Moreover the knowledge of orbital ephemeris allows the distribution of outburst events in individual systems to be studied in greater detail and provides the building blocks, and sanity checks, for the orbital models which are built to describe them. We encourage continued survey coverage of these sources, along with further high time resolution studies at targeted orbital phases in order to expand our knowledge of these extreme systems and understand their place within the HMXB family.

\end{document}